\documentclass[twocolumn]{aastex631}

\newcommand{\wise}{WISE~J062309.94$-$045624.6}
\newcommand{\askap}{ASKAP~J062309.2$-$0456227}
\usepackage{siunitx}

\begin{document}

\title{Periodic Radio Emission from the T8 Dwarf \wise} 

\correspondingauthor{Kovi Rose}
\email{kovi.rose@sydney.edu.au}

\author[0000-0002-7329-3209]{Kovi Rose}
\affiliation{Sydney Institute for Astronomy, School of Physics, The University of Sydney, New South Wales 2006, Australia}

\author[0000-0003-1575-5249]{Joshua Pritchard}
\affiliation{Sydney Institute for Astronomy, School of Physics, The University of Sydney, New South Wales 2006, Australia}
\affiliation{Australia Telescope National Facility, CSIRO, Space and Astronomy, PO Box 76, Epping, NSW 1710, Australia}
\affiliation{ARC Centre of Excellence for Gravitational Wave Discovery (OzGrav), Hawthorn, Victoria, Australia}

\author[0000-0002-2686-438X]{Tara Murphy}
\affiliation{Sydney Institute for Astronomy, School of Physics, The University of Sydney, New South Wales 2006, Australia}
\affiliation{ARC Centre of Excellence for Gravitational Wave Discovery (OzGrav), Hawthorn, Victoria, Australia}

\author[0000-0002-4079-4648]{Manisha Caleb}
\affiliation{Sydney Institute for Astronomy, School of Physics, The University of Sydney, New South Wales 2006, Australia}
\affiliation{ASTRO3D: ARC Centre of Excellence for All-sky Astrophysics in 3D, ACT 2601, Australia}

\author[0000-0003-0699-7019]{Dougal Dobie}
\affiliation{Australian Research Council Centre of Excellence for Gravitational Wave Discovery (OzGrav)}
\affiliation{Centre for Astrophysics and Supercomputing, Swinburne University of Technology, Hawthorn, Victoria 3122, Australia}

\author[0000-0002-4405-3273]{Laura Driessen}
\affiliation{Sydney Institute for Astronomy, School of Physics, The University of Sydney, New South Wales 2006, Australia}

\author[0000-0002-3846-0315]{Stefan W.~Duchesne}
\affiliation{Australia Telescope National Facility, CSIRO Space and Astronomy, PO Box 1130, Bentley WA 6102, Australia}

\author[0000-0001-6295-2881]{David~L.~Kaplan}
\affiliation{Center for Gravitation, Cosmology, and Astrophysics, Department of Physics, University of Wisconsin-Milwaukee, P.O. Box 413, Milwaukee, WI 53201, USA}

\author[0000-0002-9994-1593]{Emil Lenc}
\affiliation{Australia Telescope National Facility, CSIRO, Space and Astronomy, PO Box 76, Epping, NSW 1710, Australia}

\author[0000-0002-2066-9823]{Ziteng Wang}
\affiliation{Curtin Institute of Radio Astronomy, Curtin University, Perth, Western Australia 6845, Australia}

\begin{abstract}

We present the detection of rotationally modulated, circularly polarized radio emission from the T8 brown dwarf \wise\ between 0.9 and 2.0\,GHz. We detected this high proper motion ultracool dwarf with the Australian SKA Pathfinder in $1.36$\,GHz imaging data from the Rapid ASKAP Continuum Survey. We observed \wise\ to have a time and frequency averaged Stokes I flux density of $4.17\pm0.41$\,mJy beam$^{-1}$, with an absolute circular polarization fraction of $66.3 \pm 9.0\%$, and calculated a specific radio luminosity of $L_{\nu}\sim10^{14.8}$\,erg s$^{-1}$ Hz$^{-1}$. In follow-up observations with the Australian Telescope Compact Array and MeerKAT we identified a multi-peaked pulse structure, used dynamic spectra to place a lower limit of $B>0.71$\,kG on the dwarf's magnetic field, and measured a $P=1.912\pm 0.005$ $\SI{}{\hour}$ periodicity which we concluded to be due to rotational modulation. The luminosity and period we measured are comparable to those of other ultracool dwarfs observed at radio wavelengths. This implies that future megahertz to gigahertz surveys, with increased cadence and improved sensitivity, are likely to detect similar or later-type dwarfs. Our detection of \wise\  makes this dwarf the coolest and latest-type star observed to produce radio emission.

\end{abstract}

\section{Introduction} \label{sec:intro}

T-dwarfs are a subclass of low mass ($\lesssim 0.075$\,$M_{\odot}$) sub-stellar objects with effective temperatures between $450$--$1500$\,K \citep{2005ARA&A..43..195K,2012ApJ...753..156K}. T-dwarfs and other spectral types later than M7 are referred to collectively as ultracool dwarfs (UCDs). UCDs are fully convective and do not possess an intermediate tachocline shearing region like more massive stars \citep{2000ApJ...542..464C,2021AJ....162...43H}. Tachocline shearing is thought to be a key component in the $\alpha\Omega$ dynamo process that powers the magnetic fields of partially convective, higher-mass stars \citep{2005PhR...417....1B}.  Nonetheless, radio observations have resulted in strong evidence for $\sim10^3$\,G  magnetic fields in T-dwarfs \citep[e.g.,][]{2012ApJ...747L..22R, 2016ApJ...818...24K}, despite the lack of a tachocline region, requiring the operation of an alternative dynamo mechanism for UCDs \citep[e.g.,][]{2009Natur.457..167C}. Hence the investigation of magnetic field production in UCDs is important for improving our understanding of stellar evolution and dynamo theory. 

The chromospheric and coronal activity generally associated with radio bursts from earlier type stars ($\leq$M6) weakens with later spectral types. \cite{2015MNRAS.454.3977R} found that UCDs later than L4 dwarfs ($T_{\rm eff}\sim1600$\,K) cannot sustain the ionization levels necessary for the atmospheric current system that produces typical stellar radio emission. 
The standard practice, therefore, is to use the processes that drive auroral emission in solar system gas giant planets \citep[e.g.,][]{https://doi.org/10.1029/JA084iA11p06554} to model the magnetic activity for brown dwarfs in the L/T transition regime (L4--T4) and later \citep{2018haex.bookE.171W}. Much of the literature focuses primarily on these late-type UCDs as the radio pulse morphology, and possibly the astrophysical generator, are different for earlier-type ($\leq$L4) UCDs \citep{2017ApJ...846...75P}. 

The generation of strong dipole fields in UCDs is thought to be tied to their fast rotation \citep{2018ApJS..237...25K}. UCDs have projected rotational velocities of $v$ $\sin i\geq10$\,km s$^{-1}$ though their rotational periods can range from $\SI{1}{\hour}$ to upwards of $\SI{20}{\hour}$ \citep[][and references therein]{2021AJ....161..224T}. T-dwarfs on average rotate more rapidly than other UCDs, reaching projected velocities up to $v$ $\sin i\sim100$\,km s$^{-1}$  \citep{2006ApJ...647.1405Z,2021AJ....161..224T}. Rapid rotation plays a critical role in the co-rotational breakdown between  UCD magnetic fields and ionospheric plasma which produces the electrical currents responsible for generating auroral emission \citep{2001P&SS...49.1067C,2012ApJ...760...59N}. 

Electron-cyclotron maser instability (ECMI) is the dominant mechanism producing coherent emission from UCDs, including the auroral emission which is understood to be modulated by the star's rotational period \citep{2006ApJ...653..690H,2008ApJ...684..644H}.
ECMI converts the free plasma energy in the auroral region -- from the perpendicular component of the cyclotron motion around the magnetic field lines -- into circularly polarized emission \citep{1982ApJ...259..844M} at the electron cyclotron frequency: $\nu_c\ = eB/2\pi m_e c\approx 2.8\times 10^6 \,B$\,\,Hz \citep{1985ARA&A..23..169D}, where $e$ is the electron charge, $B$ is the magnetic field strength in gauss, $m_e$ is the electron mass, and $c$ is the speed of light
\citep{2018haex.bookE.171W}.
Analysis of radio emission from T-dwarfs therefore allows us to measure the strength and structure of their magnetic fields.
Auroral ECMI is rotationally modulated and thus the emission can be also be used to measure  rotational velocities.
This can be difficult to measure through Zeeman Doppler Imaging for late-type UCDs, which tend to be faster rotators than earlier-type M-dwarfs.
Studying these magnetic and rotational properties help us improve upon existing models of stellar dynamo theory as well as the evolution of giant exoplanets and late-type stars \citep[e.g.,][]{2009ApJ...699L.148S,2015ApJ...808..189W,2016ApJ...830L..27R,2017ApJ...846...75P}.  

The first radio detection of a UCD was reported by \citep{2001Natur.410..338B} who identified both quiescent and flaring radio emission from the M9-dwarf LP944$-$20. Radio observations conducted in the megahertz to gigahertz range have since resulted in detections of late-type L-dwarfs and T-dwarfs
\citep{2012ApJ...747L..22R,2015ApJ...808..189W,2018ApJS..237...25K}. \cite{2020ApJ...903L..33V} made the first radio detection of a UCD that had not been previously identified in optical or infrared. \cite{2020ApJ...903L..33V} used the Low-Frequency Array \citep[LOFAR;][]{2022A&A...659A...1S} to identify BDR~J1750$+$3809 which they then spectroscopically classified as a T6.5$\pm$1 dwarf with the near-infrared SpeX instrument on NASA's Infrared Telescope Facility \citep[IRTF;][]{2003PASP..115..362R}. The latest-type UCD detected in the radio to date is part of the T-dwarf binary discovered by \cite{2023arXiv230101003V}, which is composed of T5.5$\pm$0.5 and T7.0$\pm$0.5 dwarfs.

Targeted radio surveys of UCDs at $4$--$9$\,GHz have been conducted with the Karl G. Janksy Very Large Array \citep[VLA;][]{2011ApJ...739L...1P} and the Australian Telescope Compact Array \citep[ATCA;][]{2011MNRAS.416..832W} \citep[e.g.,][]{2006ApJ...648..629B, 2016MNRAS.457.1224L}. These surveys targeted a range of known UCDs and detected radio emission from $\lesssim10$\% of them \citep{2016ApJ...830...85R}. \cite{2016ApJ...818...24K} conducted a targeted VLA survey of UCDs that had exhibited signs of auroral activity at other wavelengths (H$\alpha$, optical, and infrared). They detected four out of the five dwarfs which had not previously been observed to produce radio emission. 

LOFAR and the Australian SKA Pathfinder \citep[ASKAP;][]{2021PASA...38....9H} have come online with high sensitivity widefield capabilities at megahertz to low-gigahertz frequencies. Stokes V (circular polarization) searches with the LOFAR Two-metre Sky Survey \citep[LoTSS;][]{2019A&A...622A...1S} and the Rapid ASKAP Continuum Survey \citep[RACS;][]{2020PASA...37...48M} have already found new UCDs which had not previously been observed to produce radio emission \citep[e.g.,][]{2021MNRAS.502.5438P,2023A&A...670A.124C}. UCD dynamo modeling by \cite{2009Natur.457..167C} predicts magnetic fields of order $10^2$\,G for Y-dwarfs and $10^3$\,G for T-dwarfs. Given the relationship between magnetic field strength and the electron cyclotron frequency, this would imply that megahertz and low-gigahertz observations are well placed to detect radio emission from late T-dwarfs and potentially even Y-dwarfs --- which remain undetected at radio wavelengths \citep{2019MNRAS.487.1994K}.

T-dwarfs are the coolest sub-stellar spectral type observed to produce radio emission and there are currently only six such systems that have been detected \citep{2017ApJ...846...75P,2020ApJ...903L..33V,2023arXiv230101003V}. 
In this paper we present the detection and analysis of an ultracool T8 dwarf found in a new untargeted gigahertz survey conducted with ASKAP. The source, \wise, is the coolest and latest-type UCD detected at radio wavelengths to date. Our detection of radio emission from \wise\ adds to the small population radio-active T-dwarfs and is the first example of multiple, high duty cycle pulses from a T-dwarf. The clear periodicity and strong spectral features of these pulses inform our understanding of the rotational and magnetospheric properties of \wise, while also providing more general insights into the astrophysical mechanism responsible for producing detectable radio emission in late-type ultracool dwarfs.

\section{Observations and Results} \label{sec:obs}

\subsection{ASKAP Detection}
We performed a search for highly circularly polarized objects in the RACS mid-band ($1.36$\,GHz) data \citep[RACS-mid;][]{racsmid}. 
The $\SI{15}{\minute}$ RACS-mid observations cover the whole sky south of Dec $=+49^{\circ}$ (covering $36,449$\,deg$^2$) with a median angular resolution of $\sim10\arcsec$ and a median sensitivity of $\sim0.15$--$0.40$\,mJy beam$^{-1}$. 
We used the circular polarization method presented by \cite{2021MNRAS.502.5438P} to identify interesting  sources. We identified \askap\ for further investigation because we did not find a nearby ($\lesssim5\arcsec$) positional cross-match to any known astronomical objects.

\askap\ was detected in RACS-mid on $\textrm{MJD } 59216$ with a time and frequency averaged Stokes I flux density of $4.17 \pm 0.41$\,mJy beam$^{-1}$, with a beam size of $9.5\arcsec \times 7.4\arcsec$, and an absolute fractional circular polarization of $f_{cp}=66.3 \pm 9.0\%$. There was no source detected within $120\arcsec$ of these coordinates in the RACS-low survey ($0.88$\,GHz) with a $5\sigma$ flux density limit of $1.81$\,mJy beam$^{-1}$; see Table \ref{tab:Radio_Observations} for details of these radio observations.

\subsection{ATCA Observations}

We observed \askap\ with ATCA in L-band ($1.1$--$3.1$\,GHz) on $\textrm{MJD } 59805$ for $\SI{6}{\hour}$ using the hybrid H168 array configuration and an additional $\SI{11}{\hour}$ on $\textrm{MJD } 59929$ in the extended 6C configuration (C3363). For both observations we used the ATCA calibrator source PKS 1934-638 as the primary flux calibrator and the  calibrator source [HB89] 0607-157 for phase calibration scans.

We reduced the data from these observations using the \texttt{Miriad} software \citep{1995ASPC...77..433S} and imaged using the Common Astronomy Software Application software \citep[\texttt{CASA};][]{2022PASP..134k4501C}. We used \texttt{tclean} with \texttt{briggs} weighting and a robust parameter of $0.5$, with the multi-scale multi-term multi-frequency synthesis (\texttt{mtmfs}) deconvolver and clean scales of $0, 4,$ and $8$ pixels. The $\textrm{MJD } 59805$ observation had limited \textit{uv} coverage due to the hybrid array configuration and the source's proximity to the celestial equatorial, resulting in an extended PSF which confused \askap\  with nearby sources. This observation was dominated by artifacts from bright off-axis emission. These artifacts are the result of bright sources in the field, as well as a resolved source $\sim135\arcsec$ from \wise.  The $\textrm{MJD } 59929$ observation produced more extensive coverage of \textit{uv} space and did not suffer from the same PSF and source confusion issues as the $\textrm{MJD } 59805$ observation. 

\begin{figure*}[!ht]
\centering
\includegraphics[width=9.0 cm]{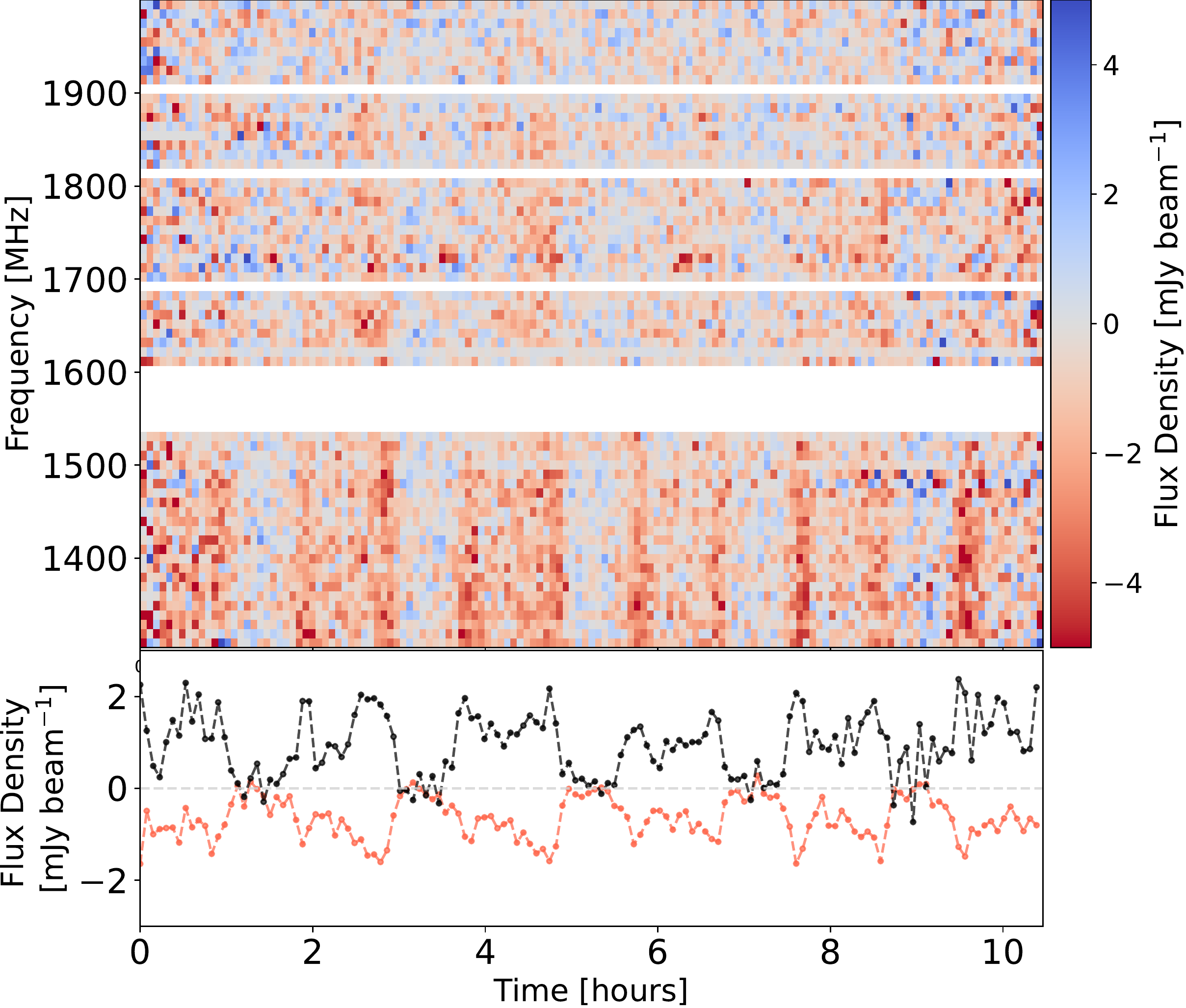}
\includegraphics[width=6.0 cm]{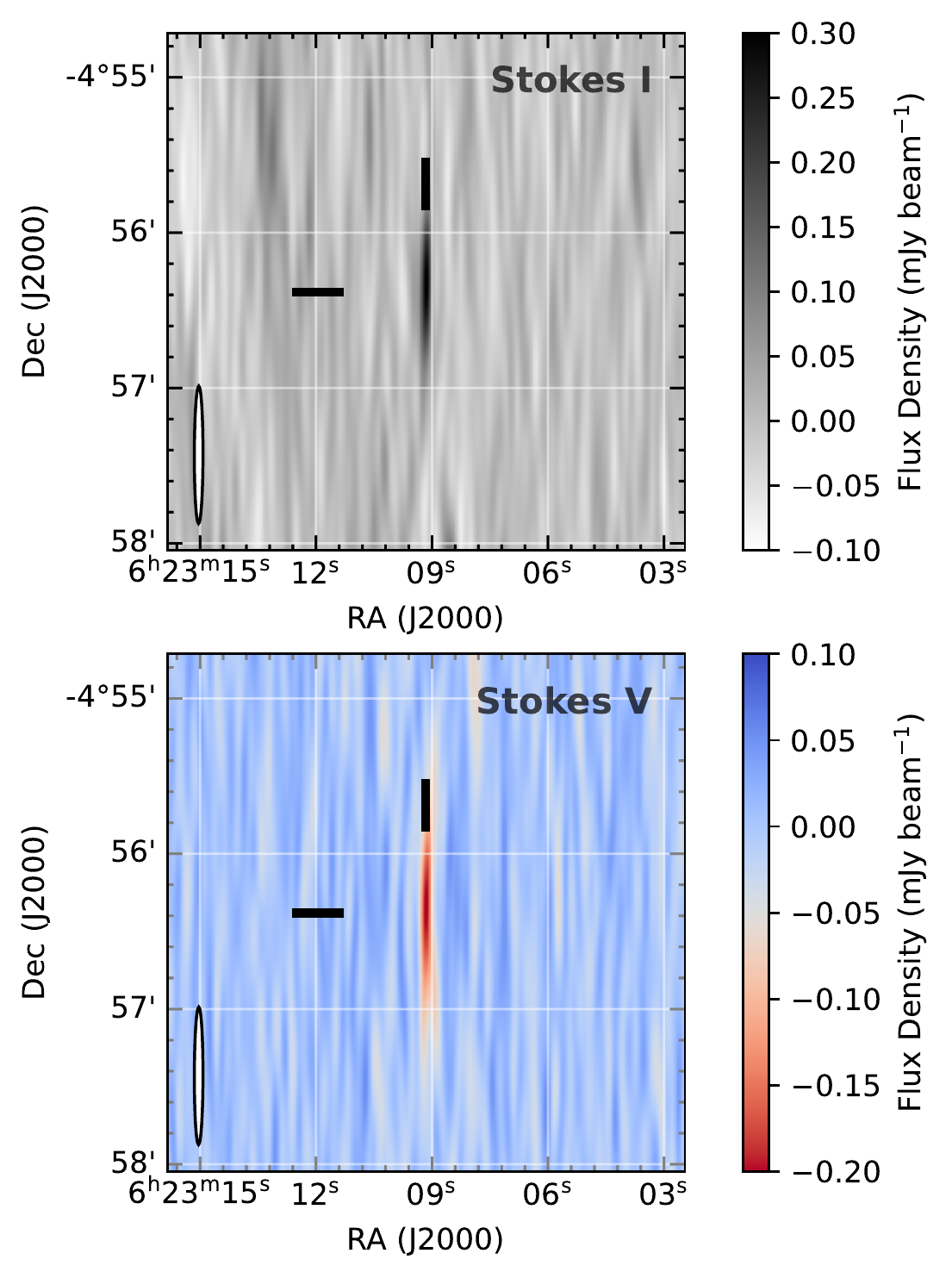}
\caption{\textit{Left:} Stokes V dynamic spectrum ($1.3$--$2.0$\,GHz) from the $\textrm{MJD } 59929$ ATCA observation. The lower panel shows the Stokes V (red) and Stokes I (black) lightcurves. We used $10$\,MHz frequency bins and $\SI{270}{\second}$ time sampling bins. Both lightcurves and the dynamic spectrum display a clear periodicity. We do not show the Stokes I dynamic spectrum as it is heavily affected by artefacts from bright off-axis emission. Horizontal gaps correspond to frequencies that were flagged due to radio frequency interference (RFI).
\textit{Right:} Stokes I and Stokes V continuum detection images from the $\textrm{MJD } 59929$ ATCA observation.}
\label{ATCA DS}
\end{figure*}

In the $\textrm{MJD } 59805$ ATCA observation we detected a time and frequency averaged Stokes I emission $0.39\pm0.06$\,mJy beam$^{-1}$, with a beam size of $10.8\arcsec \times 2.6\arcsec$, and marginal evidence of periodicity identified in the dynamic spectrum of the observation. We did not measure a clear Stokes V flux for the $\textrm{MJD } 59805$ ATCA observation. Our $\textrm{MJD } 59929$ observation produced a source detection with a time and frequency averaged Stokes I flux density of $0.30 \pm 0.03$\,mJy beam$^{-1}$, with a beam size of $53.1\arcsec \times 3.3\arcsec$, and an absolute fractional circular polarization of $f_{cp}=66.0 \pm 11.9\%$. Figure \ref{ATCA DS} shows the Stokes V dynamic spectrum as well as Stokes I and V lightcurves from the $\textrm{MJD } 59929$ ATCA observation. We see clear evidence of highly circularly polarized periodic pulsed emission.

\subsection{MeerKAT Observations}

\begin{figure*}[!ht]
\centering
\includegraphics[width=9.0 cm]{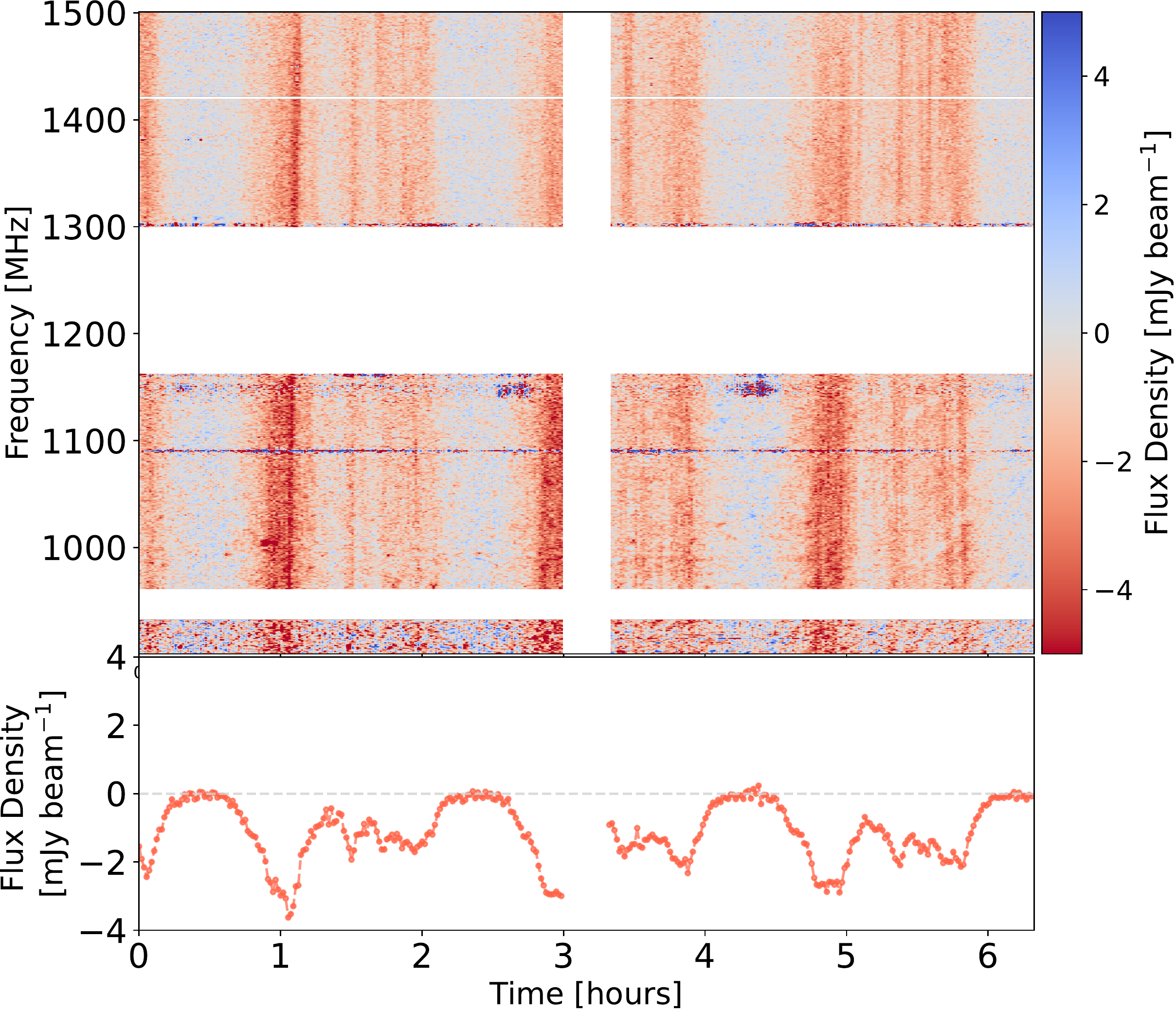}
\includegraphics[width=6.0 cm]{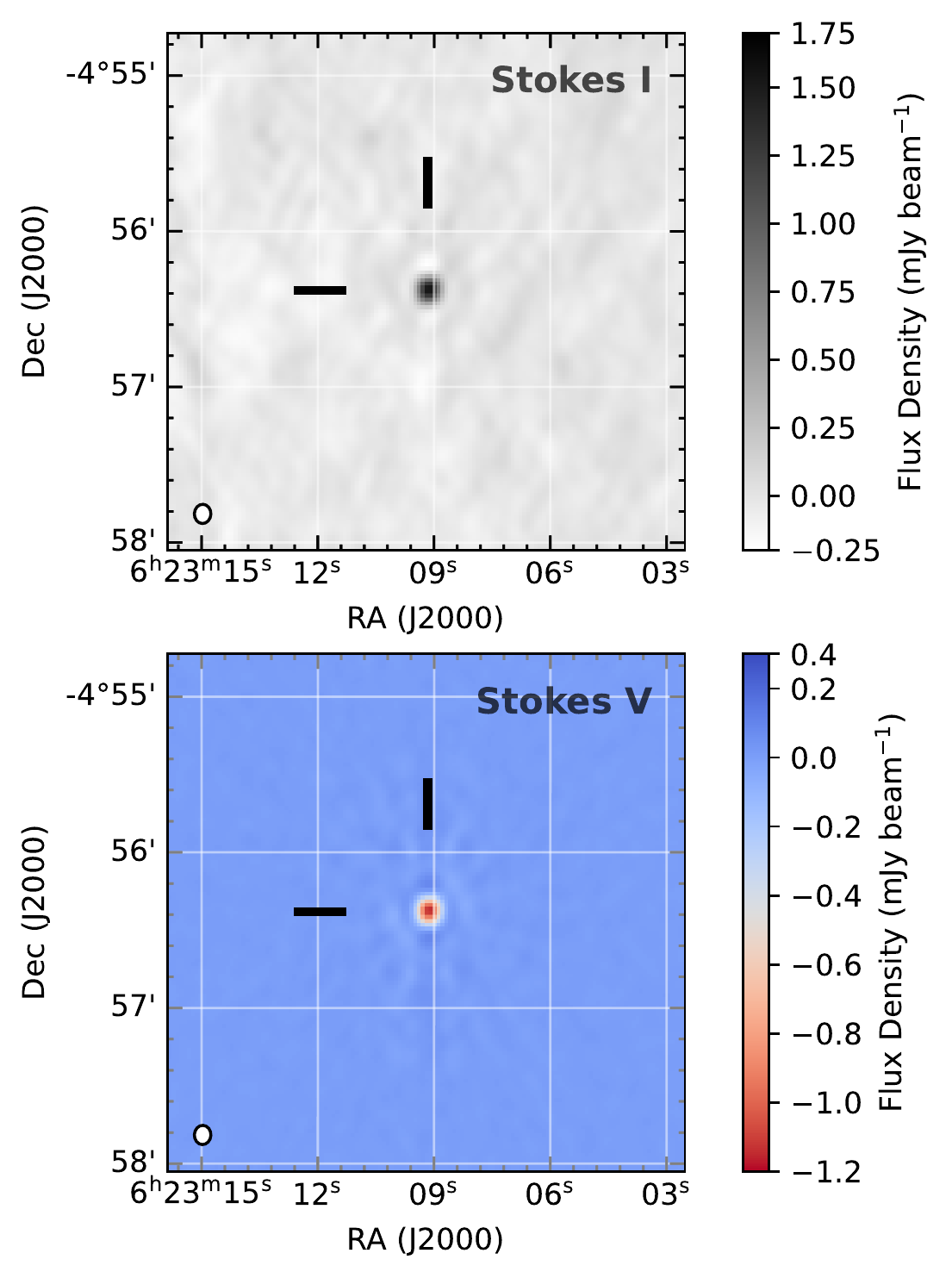}
\caption{\textit{Left:} Stokes V dynamic spectrum ($0.9$--$1.5$\,GHz) from the MeerKAT observation. The lower panel shows the Stokes V lightcurve. We used $1$\,MHz frequency bins and $\SI{64}{\second}$ time sampling bins. The lightcurve and the dynamic spectrum display periodic behavior with complex sub-pulse structure. We do not include the Stokes I dynamic spectrum or lightcurve as they are heavily affected by artefacts from bright off-axis emission. Horizontal gaps correspond to frequencies that were flagged due to radio frequency interference (RFI) and the vertical gap corresponds to a calibration scan.
\textit{Right:} Stokes I and Stokes V continuum detection images from the MeerKAT observation.}
\label{MeerKAT DS}
\end{figure*}

We observed \askap\ with MeerKAT \citep{2016mks..confE...1J} in L-band ($0.86$--$1.71$\,GHz) on $\textrm{MJD } 60030$ in the c856M4k configuration with the standard $\SI{8}{\second}$ integration time. We used the source PKS J0408-6544 as the primary flux calibrator, PKS J0521+1638 as the polarization calibrator, and the source PKS J0609-1542 for phase calibration scans. We reduced and imaged the data from this observation with the \texttt{oxkat} pipeline \citep{2020ascl.soft09003H} and used the IDIA (Inter-University Institute for Data Intensive Astronomy) \texttt{processMeerKAT}  pipeline \citep{9560276} for cross polarization calibration. We used \texttt{tclean} with \texttt{briggs} weighting and a robust parameter of $0.0$, with the multi-scale multi-term multi-frequency synthesis (\texttt{mtmfs}) deconvolver and clean scales of $0, 5, 10,$ and $15$ pixels. The MeerKAT observation produced a source detection with a time and frequency averaged  Stokes I flux density of $1.65 \pm 0.17$\,mJy beam$^{-1}$, with a beam size of $7.4\arcsec \times 6.4\arcsec$, and an absolute fractional circular polarization of $f_{cp}=73.8 \pm 10.6\%$. Figure \ref{MeerKAT DS} shows the Stokes V dynamic spectrum and lightcurve from the $\textrm{MJD } 60030$ MeerKAT observation. We see strong evidence of repeating, highly circularly polarized pulses with multiple intra-pulse peaks.

\subsection{Multi-wavelength Identification}

In RACS-mid, \askap\ has coordinates ${\rm RA_{J2000}}=06^{\rm h}23^{\rm m}09\fs28$, ${\rm Dec_{J2000}}=-04^{\circ}56'22\farcs8$ ($l=214\fd13$, $b=-8\fd53$) on $\textrm{MJD } 59216$, with uncertainties of $\pm2\arcsec$ in both RA and Dec \citep{2020PASA...37...48M,racsmid}. We extended our cross-match radius and found \wise, with an on-sky separation of $10\arcsec$ from \askap\ in the SIMBAD astronomical database \citep{2000A&AS..143....9W}. 
\wise\ has a proper motion of $\mu_{\alpha}=-0.93\pm0.01\,{\rm arcsec\,yr}^{-1}$ and $\mu_{\delta}=0.17\pm0.02\,{\rm arcsec\,yr}^{-1}$ in RA and Dec, respectively, in the CatWISE2020 catalog \citep{2021ApJS..253....8M}. Propagating the WISE object coordinates to the RACS-mid epoch resulted in an offset of $0\farcs3$ from \askap.We therefore identified \wise\ as the source of the radio emission.

\wise\ is a high proper motion T8 dwarf discovered and spectroscopically confirmed (with IRTF/SpeX) by \cite{2011ApJS..197...19K}. This UCD has an effective temperature of $T_{\rm eff}=699$\,K and is located $11.44\pm 0.38$\,pc away \citep{2019ApJS..240...19K}. \cite{2021ApJ...921...95Z} derived constraints on the age, radius, and mass of \wise\ with Sonora-Bobcat models \citep{2021ApJ...920...85M}: $\text{age}=738^{+2701}_{-592}\,{\rm Myr}$, $M=13.18^{+31.26}_{-9.44}\,{\rm M_{Jup}}$, $R=0.78^{+0.17}_{-0.13}\,{\rm R_{Jup}}$.

\wise\ was not detected in the VLA Sky Survey \citep[VLASS;][]{2020PASP..132c5001L} epochs 1.1 and 2.1, with $5\sigma$ limits of $0.57$\,mJy and $0.82$\,mJy respectively. 
Looking at the lower $700$\,MHz of the ATCA observing band (see Figure \ref{ATCA DS}), the pulsed emission stops at a cutoff frequency $\sim1.9$--$2.0$\,GHz. This is likely the reason for the $3$\,GHz VLA non-detections listed in Table \ref{tab:Radio_Observations}, though the lack of detection could also be due to the short duration observations in the VLASS on-the-fly observing mode.
This also explains the lower time and frequency averaged flux density measured with ATCA compared to ASKAP and MeerKAT, which have narrower bandwidths that are below the cutoff frequency.

We did not identify archival radio observations of \wise\ from ATCA, ASKAP, the Atacama Large Millimeter/submillimeter Array \citep[ALMA;][]{2009IEEEP..97.1463W}, or the National Radio Astronomy Observatory (NRAO) VLA Sky Survey \citep[NVSS;][]{1998AJ....115.1693C}. Nor were there detections with the ROentgen SATellite \citep[ROSAT;][]{1982AdSpR...2d.241T} or other X-ray telescopes. 

\startlongtable
\begin{deluxetable*}{ccccccccc}
\tablecaption{Radio Detections and $5\sigma$ Limits of \wise.
\label{tab:Radio_Observations}} 
\tablehead{
\colhead{Date} & \colhead{Obs. Length} & \colhead{$\nu$} & \colhead{Bandwidth} &\colhead{$S_{I}$} & \colhead{RMS$_{I}$} & \colhead{$S_{V}^a$}  & \colhead{RMS$_{V}$} & \colhead{Telescope} \\
\colhead{$[\textrm{MJD}]$} & \colhead{$[\textrm{hours}]$} & \colhead{$[\textrm{GHz}]$}&\colhead{$[\textrm{MHz}]$} & \colhead{$[\textrm{mJy beam$^{-1}$ }]$} & \colhead{$[\textrm{mJy}]$} & \colhead{$[\textrm{mJy beam$^{-1}$ }]$} & \colhead{$[\textrm{mJy}]$}
}
\startdata
$58087.481$ & OTF & $3.00$ & $2000$& $<0.57$ & $0.11$ & ... & ... & VLA \\ [0.1ex]
$58602.113$ & 0.25 & $0.88$ & $288$& $<1.80$ & $0.36$ & $<1.71$ & $0.57$ & ASKAP \\ [0.1ex]
$59131.620$ & OTF & $3.00$ & $2000$& $<0.82$ & $0.16$ & ... & ... & VLA \\ [0.1ex]
$59216.370$ & 0.25 & $1.36$ & $288$& $4.17 \pm 0.41$ & $0.21$ &  $-2.77 \pm 0.26$ & $0.14$ & ASKAP \\ [0.1ex]
$59805.842$ & 6 & $2.11$ & $2048$ & $0.39\pm 0.06$ & $0.04$ & ... & $...$ & ATCA (H168) \\ [0.1ex]
$59929.397$ & 11 & $2.11$ & $2048$ & $0.30 \pm 0.03$ & $0.02$ & $-0.20 \pm 0.03$ & $0.07$ & ATCA (6C) \\ [0.1ex]
$60030.531$ & 6.5 & $1.28$ & $770$ & $1.65 \pm 0.17$ & $0.03$ & $-1.220 \pm  0.122$ & $0.004$ & MeerKAT \\ [0.1ex]
\enddata
\tablecomments{The ASKAP non-detection is from RACS-low (SBID 8592) while the ASKAP detection, originally named \askap, is from RACS-mid (SBID 21060). VLA limits are from the VLASS epochs 1.1 and 2.1, which were observed on-the-fly (OTF). $\nu$ is the central observing frequency in gigahertz and $S_{I}$, $S_{V}$ are the peak Stokes I and Stokes V continuum flux densities at that frequency. The flux density errors  are the quadrature addition of the fitted error, RMS, and brightness uncertainty scaling --- $6$\% for ASKAP and $10$\% for MeerKAT and ATCA.
\\$^a$We follow the IAU/IEEE convention \citep{Hamaker1996} where $S_{V}>0$ corresponds to right-handed circular polarization and $S_{V}<0$ to left-handed circular polarization.}
\end{deluxetable*}
\vspace{-4mm}

\section{Analysis}
\subsection{Pulse Periodicity}
\begin{figure*}[!ht]

\centering
\includegraphics[width=7.5cm]{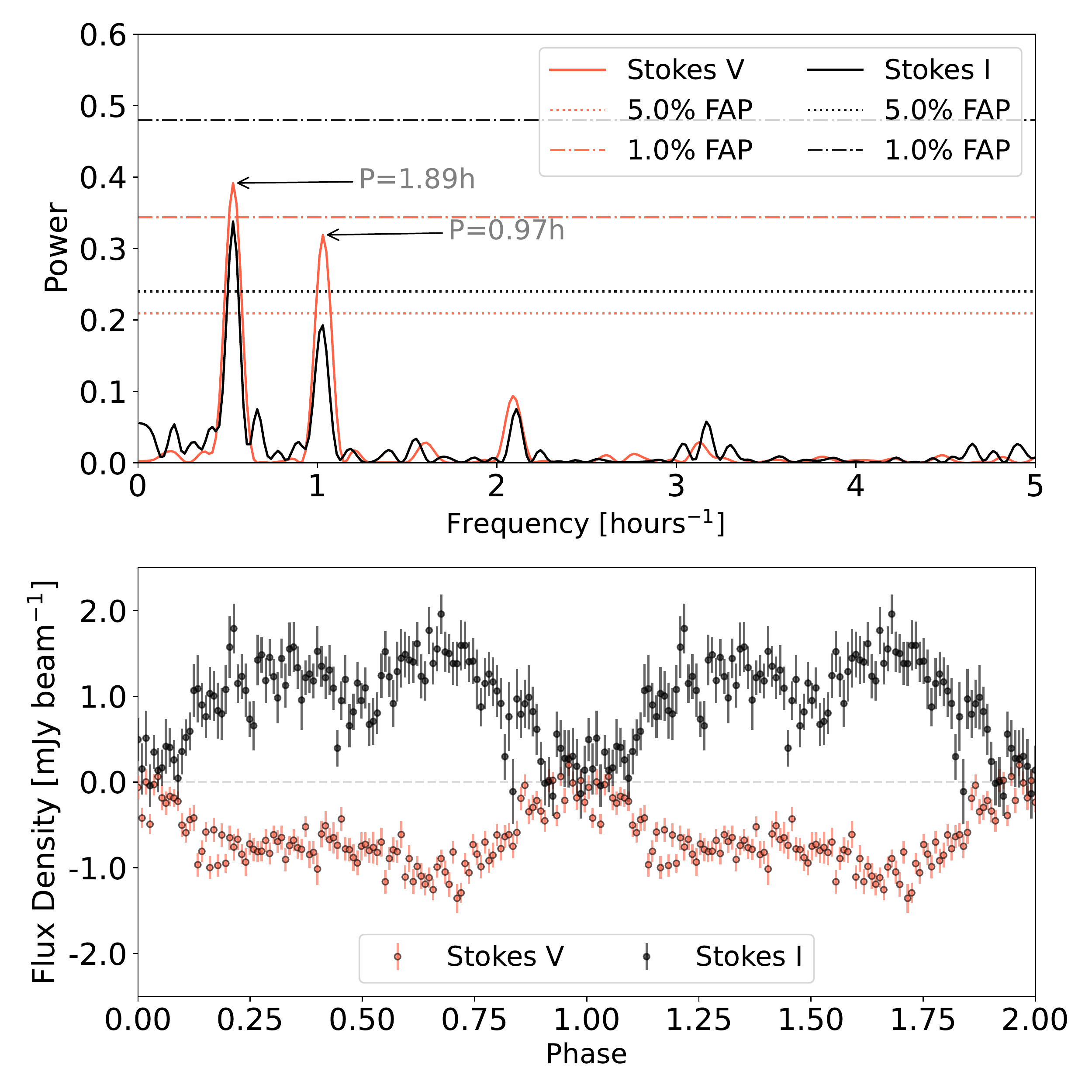}
\includegraphics[width=7.5cm]{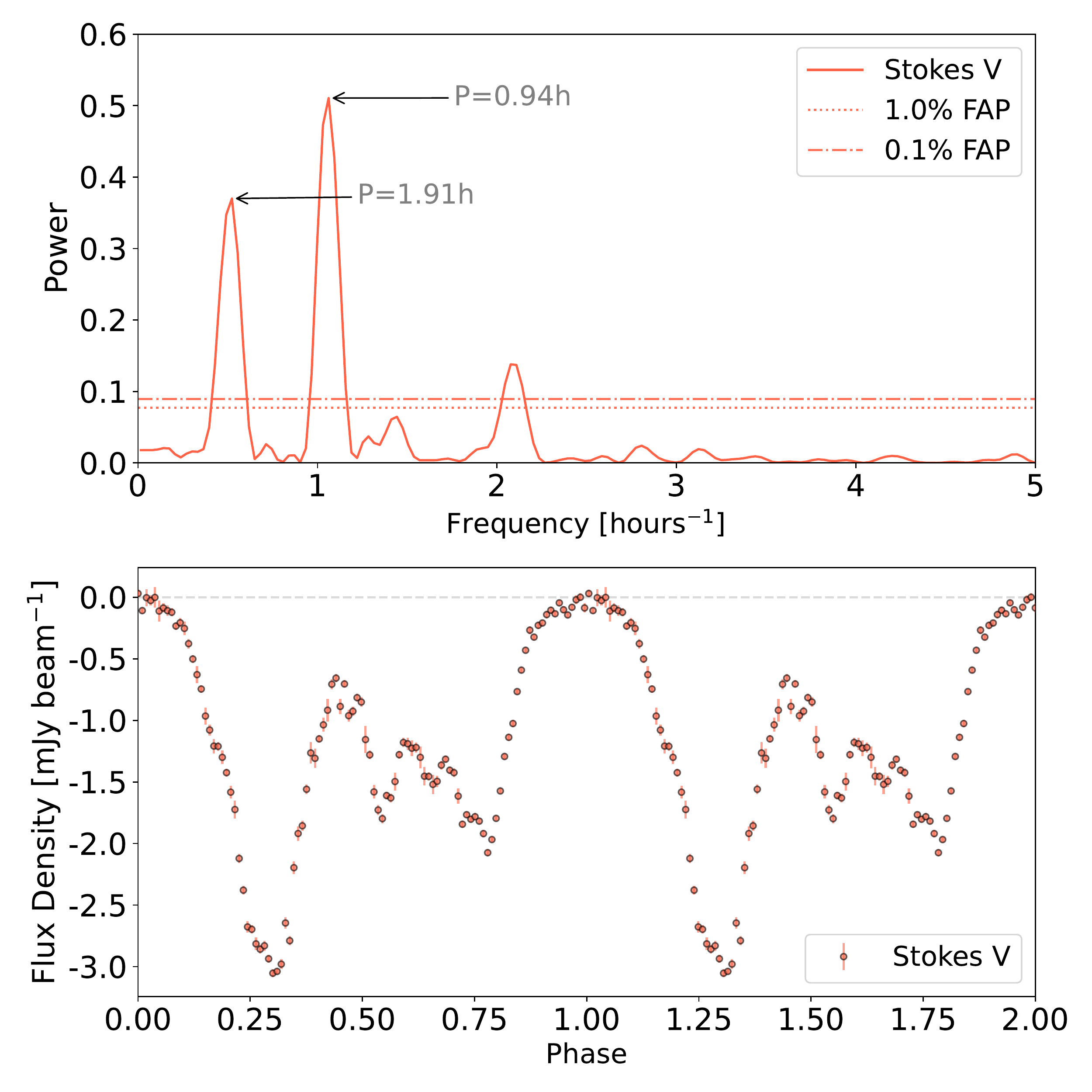}
\caption{\textit{Left: }Lomb-Scargle periodogram of the $\SI{270}{\second}$ time-sampled Stokes I and Stokes V data from the ATCA $\textrm{MJD } 59929$ observation. We plot the Stokes I (black) and Stokes V (red) power levels corresponding to the $5$\% (dotted) and $1$\% (dot-dashed) false alarm probabilities for reference (\textit{top}). The \textit{lower} panel shows the $\SI{165}{\second}$ time-sampled Stokes I and Stokes V ATCA lightcurves phase folded to $P=\SI{1.89}{\hour}$. For both panels we use the lower $700$\,MHz of the ATCA observing band with $10$\,MHz frequency binning.
\textit{Right: }Lomb-Scargle periodogram of the $\SI{64}{\second}$ time-sampled Stokes V data from the MeerKAT observation. We plot the Stokes V (red) power levels corresponding to the $0.1$\% (dot-dashed) and $1$\% (dotted) false alarm probabilities for reference (\textit{top}). The \textit{lower} panel shows the $\SI{105}{\second}$ time-sampled Stokes V MeerKAT lightcurves phase folded to $P=\SI{1.91}{\hour}$. For both panels we use the $0.9$--$1.5$\,GHz part of the MeerKAT observing band with $1$\,MHz frequency binning.}
\label{LS Plot}

\end{figure*}

We used a Lomb-Scargle periodogram to identify the dominant period in the Stokes V emission from the ATCA $\textrm{MJD } 59929$ observation. The pulsed emission has a double-peaked structure (as seen in the lower left panel of Figure \ref{ATCA DS}), which repeats with a period of $P=1.889\pm0.018$ $\SI{}{\hour}$ at a false alarm probability of $<1\%$ \citep{2018ApJS..236...16V}. Peaks within the pulses are separated by $\SI{0.971}{\hour}$, approximately half the pulse period, which appears as a less prominent peak in the Lomb-Scargle power spectrum. We used similar Lomb-Scargle analysis to find a pulse perodicity of $P=1.912\pm 0.005$ $\SI{}{\hour}$, at a false alarm probability of $<1\%$ \citep{2018ApJS..236...16V}, in the MeerKAT data. The pulsed emission has a complex pulse profile with multiple peaks (as seen in the lower left panel of Figure \ref{MeerKAT DS}). The leading and trailing peaks of the pulse are separated by $\SI{0.942}{\hour}$, approximately half the pulse period. False alarm probabilities were calculated using bootstrap resampling \citep{2018ApJS..236...16V} of the $\SI{270}{\second}$ (ATCA) and $\SI{64}{\second}$ (MeerKAT) time-averaged data. We calculated the period uncertainties using the method presented in equation 52 of \cite{2018ApJS..236...16V}, which assumes that the periods are physical and not aliases.
The pulse periodicities observed with ATCA and MeerKAT (see lower panel of Figure \ref{MeerKAT DS}) are consistent with a $P\sim\SI{1.9}{\hour}$ period.
Figure \ref{LS Plot} shows the Lomb-Scargle periodograms and
phase-folded lightcurves from the ATCA $\textrm{MJD } 59929$ Stokes I and Stokes V data and from the MeerKAT Stokes V data. Two phase cycles are shown for clarity and the folded lightcurves are cumulatively binned. Both plots in Figure \ref{LS Plot} show that $P\sim\SI{1.9}{\hour}$ is the dominant period in both the Stokes I and Stokes V emission. A less prominent peak is visible in both periodograms at $\SI{0.48}{\hour}$. This peak appears to be a harmonic corresponding to a quarter of the period.

\subsection{Radio Luminosity}

Based on modeling of \wise\ by \cite{2021ApJ...921...95Z} we assume an upper limit of $0.95\,{\rm R_{Jup}}$ for the emission region. Using the RACS-mid flux density we obtain a lower limit for the brightness temperature of $T_b>2.6\times10^{12}$\,K, which requires a coherent emission mechanism.

Using a distance of $11.44\pm 0.37$\,pc \citep{2019ApJS..240...19K}, the measured RACS-mid flux density corresponds to an isotropic radio luminosity of $L_{\nu}\sim10^{14.8}$ \,erg s$^{-1}$ Hz$^{-1}$. The peak radio luminosities of UCDs presented by \cite{2017ApJ...846...75P} range from $10^{13}$--$10^{15.5}$\,erg s$^{-1}$ Hz$^{-1}$. The isotropic radio luminosity we calculate for \wise\ is comparable to the most luminous UCDs of similar spectral types (see Table \ref{tab:UCD Radio Literature}).

\startlongtable
\begin{deluxetable*}{lccccccc}
\tablecaption{Known Late-type UCDs with Published Radio Detections.
\label{tab:UCD Radio Literature}} 

\tablehead{
\colhead{Name} & \colhead{SpT} & \colhead{$T_{\rm eff}$} & \colhead{$\log_{10}(L_{\nu,\rm avg})$} & \colhead{$\log_{10}(L_{\nu,\rm peak})$} & Band & \colhead{Ref.}& \\
 & & \colhead{$[\textrm{K}]$}&\colhead{[\,erg s$^{-1}$ Hz$^{-1}$]} & \colhead{[\,erg s$^{-1}$ Hz$^{-1}$]} & [GHz] & &
}
\startdata
2MASS J04234858$–$0414035 & L6 & $1483 \pm 113$ & $12.8$& $13.7$ & $4.0$--$12.0$ &  \cite{2016ApJ...818...24K,2018ApJS..237...25K} \\ [0.1ex]
2MASS 10430758$+$2225236 & L8 & $1336 \pm 113$ & $12.7$& $13.4$  & $4.0$--$12.0$ &  \cite{2016ApJ...818...24K,2018ApJS..237...25K} \\ [0.1ex]
SIMP J013656.5$+$093347.3 & T2.5 & $1089 \pm 62$ & $12.2$& $13.0$  & $4.0$--$12.0$ & \cite{2016ApJ...818...24K,2018ApJS..237...25K}\\ [0.1ex]
WISEP J112254.73$+$255021.5$^a$ & T6 & $943 \pm 113$ & ... & $14.9$ & $4.0$--$7.0$ &  \cite{2016ApJ...821L..21R} \\ [0.1ex]
2MASS J10475385$+$2124234$^b$ & T6.5 & $880 \pm 76$ & $12.1$ & $13.2$  & $4.0$--$18.0$ & \cite{2015ApJ...808..189W}\\ [0.1ex]
2MASS J12373919$+$6526148 & T6.5 & $851 \pm 74$ & $12.7$ & $13.1$  & $4.0$--$12.0$ &  \cite{2016ApJ...818...24K,2018ApJS..237...25K}\\ [0.1ex]
BDR~J1750$+$3809 & T6.5 & ... & $15.0$ & ... & $0.12$--$0.17$ &  \cite{2020ApJ...903L..33V}  \\ [0.1ex]
WISEP J101905.63$+$652954.2$^c$ & T7+T5.5 & ... & ... & $14.0$  & $0.12$--$0.17$ &  \cite{2023arXiv230101003V}\\ [0.1ex]
\wise\ & T8 & $699$ & $ ... $ & $14.8$ & $0.9$--$2.0$ &  This Work \\ [0.1ex]
\enddata
\tablecomments{We include average and peak radio luminosities, arranged by spectral type (SpT), for dwarfs at the L/T transition (L4--T4) and cooler from \cite{2017ApJ...846...75P,2020ApJ...903L..33V,2023arXiv230101003V}, and references therein. We list the observing bands the UCDs have been detected.
\\$^a$ WISEP J112254.73$+$255021.5 was first detected in radio by \cite{2016ApJ...821L..21R} in the $4.0$--$6.0$\,GHz band and then by \cite{2017ApJ...834..117W} in the $5.0$--$7.0$\,GHz band.
\\$^b$ 2MASS J10475385$+$2124234 was detected in the $5.0$--$7.0$\,GHz and $9.0$--$11.0$\,GHz bands by \cite{2015ApJ...808..189W}, $4.0$--$6.0$\,GHz by \cite{2016ApJ...830...85R}, and $12.0$--$18.0$\,GHz by \cite{2018ApJS..237...25K}. 
\\$^c$ The latest-type radio-emitting UCD prior to this work is the T7+T5.5 binary detected by \cite{2023arXiv230101003V} at $144$\,MHz.} 
\end{deluxetable*}

\section{Discussion}
\subsection{Model-dependent Source Parameters} 

The detected emission from \wise\ has a brightness temperature beyond the $10^{12}$\,K coherent limit, implying that it is generated either by plasma radiation or by ECMI. ECMI is the primary source of coherent emission in UCDs which is both highly circularly polarized and rotationally modulated   
\citep{2006ApJ...653..690H,2008ApJ...684..644H}. Because the \wise\ emission is strongly circularly polarized and periodic, we favor the ECMI interpretation. This implies that the observed periodicity corresponds to the dwarf's rotation period.

Using a conservative emission upper cutoff of $2.0$\,GHz from the $\textrm{MJD } 59929$ ATCA observation, the minimum magnetic field strength in the emission region is $B=2.0/2.8 \simeq 0.71$\,kG \citep{1985ARA&A..23..169D}.
However, the ECMI mechanism could be operating at the first harmonic of the cyclotron frequency rather than the fundamental, in which case the magnetic field lower limit would instead be $B\simeq 0.35$\,kG. The upper cutoff for ECMI emission is typically interpreted to represent the magnetic field strength at the base of the emission region, where the plasma density becomes sufficient to drive the local plasma frequency $\nu_p$ above the cyclotron frequency $\nu_c$, and in principle provides a lower limit to the maximum stellar magnetic field strength at the photosphere. Because we detect radio emission up to
$2.0$\,GHz in the $\textrm{MJD } 59929$ ATCA observation, we used the electron plasma frequency relation: $\nu_p\approx9\times10^3 n_e^{1/2}$\,Hz \citep{1985ARA&A..23..169D} to calculate an upper limit of $n_e<5\times10^{10}$\,cm$^{-3}$ on the local electron density at the base of the emission region.
The spectral cutoff may also be due to the plasma frequency, which is dependent on the electron density, reaching the order of the cyclotron frequency at a higher altitude void in the dwarf's magnetosphere. This would cut off the emission at a lower magnetic field strength and the true maximum magnetic field strength would be larger.

Using the upper limit radius $r=0.95\,{\rm R_{Jup}}$ \citep{2021ApJ...921...95Z} we can use the $P=\SI{1.9}{\hour}$ period to constrain the rotational velocity to $v$ $\sin i> 63$\,km s$^{-1}$. This projected rotational velocity is relatively high for a UCD but is still slower than the fastest known rotating T-dwarfs found by \cite{2021AJ....161..224T} and \cite{2022ApJ...924...68V}. The rotational period is reasonable as it is above the $\SI{1}{\hour}$ empirical lower limit for UCD rotation periods \citep{2021AJ....161..224T} and well above the break-up period lower limit $P_{\rm break-up}=0.58^{+0.38}_{-0.22}\,\SI{}{\hour}$. We calculated this break-up limit by taking  the parallax $\varpi=86.5 \pm 1.7$\,mas and the Starfish-based values for the surface gravity $\log g=4.70^{+0.47}_{-0.42}$\,dex and solid angle $\log\Omega=-19.610^{+0.167}_{-0.156}$\,dex from \cite{2021ApJ...921...95Z}. We then computed the breakup period as $P_{\rm break-up}=\sqrt{4\pi^2\Omega^{1/2}/g\varpi G }$ by using a Monte Carlo estimate for the uncertainty, drawing the values for $\log\Omega$, $\log g$, and $\varpi$ from independent Gaussians and computing the distribution in values of $P_{\rm break-up}$ that resulted.

\subsection{Source of ECMI Emission}\label{sec:ecmi-beaming}
The pulse profile detected in both ATCA and MeerKAT observations of \wise\ has a dominant
period of $\SI{1.9}{\hour}$ which we interpret as the stellar rotation period, and a duty cycle of $\sim70\%$. 
Each pulse profile features a multi-peaked structure with strong leading and trailing pulses separated by a phase
of $\sim 0.5$, multiple weaker pulses inbetween, and an apparent ``off" period for the remaining
$\sim 0.5$ phase. ECMI radio emission is beamed into a hollow cone aligned with the local
magnetic field and with half-angle that is dependent upon the nature of the instability in the
electron velocity distribution, typically near perpendicular to the magnetic field
\citep{Treumann2006}. Emission from an extended region of the stellar magnetosphere such as an
auroral oval is a superposition of individual ECMI sources with complex, overlapping beaming
geometry.

The $\sim 0.5$ phase separation of the strongest pulse peaks in our data can then be explained
by the leading and trailing edges of an extended ECMI source analogous to Jupiter's main
auroral oval \citep[e.g.,][]{2001P&SS...49.1067C}. This is the expected ECMI source distribution for aurorae that are powered by co-rotation breakdown between the stellar magnetosphere and
ionosphere, which is a plausible scenario given the rapid $\SI{1.9}{\hour}$ rotation period.
Alternatively, a localized ECMI source with perpendicular beaming angle may produce strong
leading and trailing pulses separated by $\sim 0.5$ phase \citep[e.g.,][]{2007ApJ...663L..25H}, and extension of the source over a small range of stellar longitudes can then explain the weaker
intermediate pulses as separate ECMI cones are swept into the line of sight
\citep[e.g.,][]{Bastian2022}.
The pulse duty cycle of $\sim70\%$ is much larger than most previously detected 
UCD auroral pulses, with only a few similar examples such as the L3.5 dwarf 2M~J0036+1821 
\citep{2008ApJ...684..644H} and the T6 dwarf WISE~J1122+2550 \citep{2017ApJ...834..117W} with duty cycles of 
$\sim 30\%$ and $\sim 50\%$ respectively, which may be explained by unresolved sub-pulse structure
as shown in Figure \ref{MeerKAT DS}.
Modeling of the 
time-frequency structure of pulses in the dynamic spectra will inform the interpretation of the ECMI source
configuration and electrodynamic engine powering the aurorae, and can provide constraints on the
rotation and magnetic axis inclinations, which will be the subject of a future publication.

\subsection{Volumetric Rate Analysis}

\wise\ is the only T-dwarf detected in our untargeted RACS-mid circular polarization search, which will be presented in an upcoming publication. While the radio luminosity distribution of T-dwarf auroral pulses is still poorly constrained, to determine an order of magnitude estimate of the expected number of T-dwarf detections we assume the observed value of $L_{\nu}\sim 10^{14.8}$ \,erg s$^{-1}$ Hz$^{-1}$ is representative of the population. With a $5\sigma$ limiting flux density of $1.25$\,mJy beam$^{-1}$ in RACS-mid, this corresponds to a detection horizon of $20.54$\,pc. 
\cite{2021AJ....161...42B} used a volume-limited $25$\,pc sample of $190$ T-dwarfs to calculate a space density of $5.45 \pm 0.24 \times10^{-3}$ T-dwarfs (T0--T8) per cubic parsec.
We therefore expect a total of $180\pm 10$ T-dwarfs within the sampled $20.54$\,pc RACS-mid sensitivity horizon, accounting for the $\sim88\%$ RACS-mid sky coverage.

However not all T-dwarfs in a given volume are necessarily radio-active. \cite{2018haex.bookE.171W} suggested that the $\leq10\%$ detection fraction for UCDs in targeted radio surveys \citep{2016ApJ...830...85R} may be due to a similarly small fraction of UCDs having the necessary magnetospheric conditions to produce detectable radio emission. 
Another factor contributing to the $\leq10\%$ detection fraction is the beaming geometry \citep{2017ApJ...846...75P}, where some fraction of T-dwarfs will have rotation and magnetic axes that are unfavorably aligned for the rotationally modulated ECMI beam to cross our line of sight. The overall $\sim 10\%$ fraction of radio pulsing UCDs implies that $\sim18$ T-dwarfs within the RACS-mid sensitivity horizon are producing detectable auroral radio pulses. 

Finally, radio-pulsing T-dwarfs whose emission is favorably beamed may have been missed due to the short duration of RACS-mid observations, which are not capable of sampling the full rotational phase of these systems. Measured T-dwarf rotational periods typically range from $\sim\SI{1}{\hour}$ to $\sim\SI{15}{\hour}$ \citep[][and references therein]{2021AJ....161..224T}, while the RACS-mid observations cover each candidate radio-loud T-dwarf for $\SI{15}{\minute}$, implying a rotation phase coverage of as little as $\sim2\%$ for slowly rotating T-dwarfs to $\sim25\%$ for rapidly rotating T-dwarfs. As discussed in Section \ref{sec:ecmi-beaming}, the duty cycle of auroral ECMI emission is dependent on multiple factors including magnetic and rotation axis orientation and the distribution of auroral sources in altitude and longitude. While the distribution of these factors in the T-dwarf population are still poorly constrained, most auroral pulses detected to date have duty cycles of less than $\sim10\%$ \citep[e.g.,][]{2016ApJ...818...24K, 2017ApJ...846...75P, 2018haex.bookE.171W}. Using the average measured T-dwarf rotation period of $\sim\SI{5}{\hour}$ and assuming a typical duty cycle of $\sim10\%$, we expect a significant number of the estimated $\sim18$ detectable aurorally active T-dwarfs in RACS-mid would be undetected due to limited rotation phase coverage. While a quantitative estimate of the expected detection rates would require stronger constraints on the population parameters of auroral T-dwarf activity, we find that the expected number of detections in RACS-mid is of order unity. This is consistent with \wise\ being the only T-dwarf we found in our circular polarization search of RACS-mid.

\section{Conclusions}

This work described our methods of discovery and analysis of \wise. We detected rotationally modulated coherent emission with a periodicity of $P=\SI{1.9}{\hour}$. This detection provides additional evidence of complex magnetic field activity in late-type UCDs, and further validation for using the auroral-planet framework to model UCD radio emission. Because their observed flux is relatively faint compared to other radio sources, we see that circular polarization searches are a useful method of identifying late-type UCDs in widefield untargeted surveys. Future deep surveys such as the ASKAP Evolutionary Map of the Universe \citep[EMU;][]{2011PASA...28..215N,2021PASA...38...46N} will enable the discovery of more nearby auroral pulsing UCDs through greater coverage of their rotational phase, while the improved instantaneous sensitivity of future surveys with the low and mid-band SKA1 \citep{2019arXiv191212699B} will push the detection horizon outwards, and perhaps enable detection of the first radio-active Y-dwarf.

\vspace{10mm}

We would like to thank the anonymous referee for reviewing this manuscript and providing us with helpful and constructive feedback. The authors would like to thank James K. Leung and Yuanming Wang for their suggestions and helpful discussions. We would also like to express our gratitude towards Ben Montet for their efforts to find additional data for this work. 
The authors would like to thank SARAO staff for scheduling our DDT observation. The MeerKAT telescope is operated by the South African Radio Astronomy Observatory, which is a facility of the National Research Foundation, an agency of the Department of Science and Innovation.
This scientific work uses data obtained from Inyarrimanha Ilgari Bundara / the Murchison Radio-astronomy Observatory. We acknowledge the Wajarri Yamaji People as the Traditional Owners and native title holders of the Observatory site. CSIRO’s ASKAP radio telescope is part of the Australia Telescope National Facility (\href{https://ror.org/05qajvd42}{ATNF}). Operation of ASKAP is funded by the Australian Government with support from the National Collaborative Research Infrastructure Strategy. ASKAP uses the resources of the Pawsey Supercomputing Research Centre. Establishment of ASKAP, Inyarrimanha Ilgari Bundara, the CSIRO Murchison Radio-astronomy Observatory and the Pawsey Supercomputing Research Centre are initiatives of the Australian Government, with support from the Government of Western Australia and the Science and Industry Endowment Fund.
The Australia Telescope Compact Array is part of the \href{https://ror.org/05qajvd42}{ATNF} which is funded by the Australian Government for operation as a National Facility managed by CSIRO. We acknowledge the Gomeroi people as the Traditional Owners of the Observatory site.
This publication makes use of data products from the Wide-field Infrared Survey Explorer (WISE), which is a joint project of the University of California, Los Angeles, and the Jet Propulsion Laboratory/California Institute of Technology, funded by the National Aeronautics and Space Administration.
DK is supported by NSF grant AST-1816492. MC acknowledges support of an Australian Research Council Discovery Early Career Research Award (project number DE220100819) funded by
the Australian Government and the Australian Research Council Centre of Excellence for All Sky Astrophysics in 3 Dimensions (ASTRO
3D), through project number CE170100013.
This research has made use of the \href{cutouts.cirada.ca}{CIRADA} cutout service, operated by the Canadian Initiative for Radio Astronomy Data Analysis (CIRADA). CIRADA is funded by a grant from the Canada Foundation for Innovation 2017 Innovation Fund (Project 35999), as well as by the Provinces of Ontario, British Columbia, Alberta, Manitoba and Quebec, in collaboration with the National Research Council of Canada, the US National Radio Astronomy Observatory and Australia’s Commonwealth Scientific and Industrial Research Organisation.
This research has made use of the VizieR catalogue access tool, \href{10.26093/cds/vizier}{CDS}, Strasbourg, France. The original description
of the VizieR service was published in \citet{2000A&AS..143...23O}.
This research has made use of the SIMBAD database,
operated at CDS, Strasbourg, France \citep{2000A&AS..143....9W}.
This research has made use of NASA’s Astrophysics Data System (\href{https://ui.adsabs.harvard.edu/}{ADS}) Bibliographic Services.
This research made use of \href{http://www.astropy.org}{Astropy}, a community-developed core Python package for Astronomy \citep{2013A&A...558A..33A,2018AJ....156..123A}.

\vspace{5mm}
\facilities{ASKAP, ATCA, MeerKAT}

\software{Numpy \citep{2020Natur.585..357H} , Matplotlib \citep{thomas_a_caswell_2023_7697899}, Pandas \citep{the_pandas_development_team_2023_7979740}, Scipy \citep{2020SciPy-NMeth}, Astropy \citep{2018AJ....156..123A}, uncertainties, \citep{uncertainties}, Miriad \citep{1995ASPC...77..433S}, CASA \citep{2022PASP..134k4501C}, oxkat \citep{2020ascl.soft09003H}, processMeerKAT \citep{9560276}}

\bibliography{main.bib}
\bibliographystyle{aasjournal}

\end{document}